\documentclass[10pt,twocolumn,letterpaper]{article}

\usepackage{wacv}              

\usepackage{graphicx}
\usepackage{amsmath}
\usepackage{amssymb}
\usepackage[accsupp]{axessibility}  
\usepackage{xcolor}
\usepackage[utf8]{inputenc} 
\usepackage[T1]{fontenc}    
\usepackage{url}            
\usepackage{booktabs}       
\usepackage{amsfonts}       
\usepackage{nicefrac}       
\usepackage{microtype}      
\usepackage{xcolor}         
\usepackage{caption}
\usepackage{subcaption}
\usepackage{float}
\usepackage{fancyhdr}
\usepackage{numprint}      
\usepackage{nicematrix}
\usepackage[pagebackref,breaklinks,colorlinks]{hyperref}

\usepackage[capitalize]{cleveref}
\crefname{section}{Sec.}{Secs.}
\Crefname{section}{Section}{Sections}
\Crefname{table}{Table}{Tables}
\crefname{table}{Tab.}{Tabs.}

\usepackage{algorithm}
\usepackage{algpseudocode}
\usepackage{booktabs}
\usepackage{array}
\usepackage{makecell}
\usepackage{multirow}

\newcommand{\bftab}{\fontseries{b}\selectfont}

\usepackage{lipsum}

\def\wacvPaperID{2687} 
\def\confName{WACV}
\def\confYear{2024}

\begin{document}

\title{Improved Topological Preservation in 3D Axon Segmentation and Centerline Detection using Geometric Assessment-driven Topological Smoothing (GATS)}

\author{Nina I. Shamsi \\
\and
Alex S. Xu \\
\and
Lars A. Gjesteby \\
\and
Laura J. Brattain \\
\and
MIT Lincoln Laboratory, Lexington, MA 02421, USA \\
{\tt\small n.shamsi@ece.neu.edu}, {\tt\small alecx@umich.edu} \\
{\tt\small \{lars.gjesteby, brattainl\}@ll.mit.edu}\thanks{\copyright 2023 Massachusetts Institute of Technology}
}

\maketitle
\begin{abstract}
   Automated axon tracing via fully supervised learning requires large amounts of 3D brain imagery, which is time consuming and laborious to obtain. It also requires expertise. Thus, there is a need for more efficient segmentation and centerline detection techniques to use in conjunction with automated annotation tools. Topology-preserving methods ensure that segmented components maintain geometric connectivity, which is especially meaningful for applications where volumetric data is used, and these methods often make use of morphological thinning algorithms as the thinned outputs can be useful for both segmentation and centerline detection of curvilinear structures. Current morphological thinning approaches used in conjunction with topology-preserving methods are prone to over-thinning and require manual configuration of hyperparameters. 
   
   We propose an automated approach for morphological smoothing using geometric assessment of the radius of tubular structures in brain microscopy volumes, and apply average pooling to prevent over-thinning. We use this approach to formulate a loss function, which we call Geometric Assessment-driven Topological Smoothing loss, or GATS. Our approach increased segmentation and centerline detection evaluation metrics by 2\%-5\% across multiple datasets, and improved the Betti error rates by 9\%. Our ablation study showed that geometric assessment of tubular structures achieved higher segmentation and centerline detection scores, and using average pooling for morphological smoothing in place of thinning algorithms reduced the Betti errors. We observed increased topological preservation during automated annotation of 3D axons volumes from models trained with GATS.
\end{abstract}

\section{Introduction}
\label{sec:intro}
Curvilinear structures are line-like objects with differences in pixel intensities relative to neighboring pixels \cite{BIBILONI2016949}; a line is a 1-dimensional manifold, though a curvilinear structure may not necessarily be 1-dimensional \cite{multiscale}. Curvilinear structure segmentation is segmenting binary masks of curvilinear structures \cite{BIBILONI2016949}. Methods that ensure maximizing geometric connectivity of segmented curvilinear structures are said to be topology-preserving \cite{cldice, JFTN, hu2022structureaware}, and are used for performing topologically accurate segmentation. A variety of domains and applications require topologically accurate segmentation: 3D axon tracing and centerline detection \cite{pollack2022axon}, retinal vessel segmentation \cite{vessel}, and airway tree reconstruction \cite{GRELARD201789, zhang2022differentiable}, to name a few. 

Skeletonization-based approaches for curvilinear structure segmentation are useful for both segmentation and centerline detection \cite{cldice, JFTN}. Morphological thinning algorithms for volumetric data have previously utilized neighborhood lookup tables \cite{post2016fast} or directional sub-iteration \cite{xie2003topology}, but such algorithms can be computationally expensive \cite{nemeth2011thinning, wagner2020real}. The centerline Dice (clDice) loss function is an example of a topology-preserving method that has shown to be performant for curvilinear structure segmentation and centerline detection, especially with respect to model training time, through a process called soft-skeletonization \cite{cldice}. However, parameterizing soft-skeletonization is currently done via educated guesswork \cite{cldice, pollack2022axon, rouge2023cascaded}. 

This paper presents a novel approach to determine the number of iterations to perform thinning or smoothing operations using the mean pixel radius of axon segments from N random slices of volumetric data. We use this method to automatically parameterize morphological thinning or smoothing algorithms. Moreover, we observed that certain affine rotation transformations would degrade the performance of models which utilized skeletonization, and so we modified our approach to use morphological smoothing via average pooling operations instead of using morphological thinning via min-, max- pooling. We show that our approach, Geometric Assessment-driven Topological Smoothing (GATS), increased segmentation and centerline detection metrics by $2\%-5\%$ across multiple datasets with 3D axon imagery. Moreover, topology-preserving methods can be used for automated annotation of unannotated volumes of brain imagery \cite{pollack2022axon, klinghoffer2020selfsupervised}. The present work shows an application of GATS for automated annotation of 3D brain imagery. As far as we know, our work is the first to present a topology-preserving morphological smoothing approach for curvilinear structure segmentation and centerline detection in 3D brain imagery. Therefore, the contributions of this work are as follow:
\begin{enumerate}
    \item a morphological smoothing method for curvilinear structures using the mean pixel radius of tubular structures across N random slices from a volumetric input and average pooling operations.
    \item automated annotation of axon segments in 3D brain imagery using our approach, which prevents over-thinning of axons and thus promotes topological preservation in automatically annotated volumes.
\end{enumerate}

\section{Related Literature}
\label{sec:rel}
\subsection{Topology-Preserving Losses for Curvilinear Structure Segmentation}
Presently the literature for curvilinear structure segmentation and centerline detection focuses on thinning algorithms, so we compare our approach to thinning and non-thinning methods. However, morphological smoothing has been utilized for 3D images \cite{nemeth2011thinning}, and biomedical images previously \cite{wyburd2021teds, tu2021topology, meirovitch2016multi}. Our research aims to maximize topology-preservation of segmented axons from 3D volumes as axons can present with various lengths and a single axon can span across an entire input volume. Moreover, as we are interested in automated annotation of unannotated brain imagery, we seek a method that can precisely and accurately detect an axon's centerline.

\begin{algorithm}
  \caption{Soft Skeletonization}\label{skelorig}
  \begin{algorithmic}[1]
    \Procedure{SKEL}{$I, k$}\Comment{k is set manually}
      \State $I' \gets maxpool(minpool(I))$
      \State $S \gets ReLU(I - I')$
      \For{\texttt{i to k}}
        \State $I \gets minpool(I)$
        \State $I' \gets maxpool(minpool(I))$
        \State $Delta \gets ReLU(I-I')$
        \State $S \gets S + ReLU(Delta - S \circ Delta) $
      \EndFor
      \State \textbf{return} $S$
    \EndProcedure
  \end{algorithmic}
\end{algorithm}

\subsubsection{Centerline Dice (clDice)} clDice was developed to measure the topology-preservation of tubular or curvilinear structure segmentation because other measures of segmentation quality, e.g. Dice, would report the same segmentation quality for models with comparably different performance for segmenting an input volume with both small/fine and large/coarse tubular structures \cite{cldice}, i.e., the degree of a model's segmentation quality was not reflected in the scoring metric. The clDice metric reflects the degree of topology preserved following segmentation by comparing the intersection of masks and skeletons using measures of topological precision (Eq. \ref{eq1}) and topological sensitivity (Eq. \ref{eq2}), where the harmonic mean of Eq. \ref{eq1} and Eq. \ref{eq2} results in a clDice score (Eq. \ref{eq3}); a higher score indicates a higher degree of structural connectivity. \emph{$S_P$} is the predicted skeleton, and \emph{$V_L$} is the ground truth volume; similarly,  \emph{$S_L$} is the ground truth skeleton, and  \emph{$V_P$} is the predicted segmentation mask. The process of computing clDice is an algorithm called soft-clDice. 

\begin{equation} \label{eq1}
Tprec(S_P, V_L) = \frac{|S_P\cap\ V_L|}{|S_P|}
\end{equation}

\begin{equation} \label{eq2}
Tsens(S_L, V_P) = \frac{|S_L\cap\ V_P|}{|S_L|}
\end{equation}

Skeletonization is the process of iteratively removing foreground pixels in a binary image till a construct remains that maintains the extent and connectivity of the original object \cite{chiang1992euclidean}. The soft skeletonization algorithm (Algorithm \ref{skelorig}), which produces the skeletons used to compute topological precision and topological sensitivity, performs morphological thinning of a curvilinear structure by applying iterative min- and max- pooling using a data-dependent hyperparameter \emph{k}. The clDice paper (2021) notes that the \emph{k} hyperparameter must be greater than the pixel radius of the largest observed tubular structure in a given dataset.

\begin{equation} \label{eq3}
clDice(V_P, V_L) = 2 \times \frac{Tprec(S_P, V_L) \times Tsens(S_L, V_P)}{Tprec(S_P, V_L) + Tsens(S_L, V_P)} 
\end{equation}

Soft-skeletonization, via soft-clDice, produces a differentiable result that can be used for neural network training, which is important because usually morphological thinning is a discrete operation and produces a non-differentiable result \cite{cldice, zhang2022differentiable}. clDice provides a balance for preserving topology while learning segmentation of curvilinear structures by setting $\alpha \in [0, 0.5]$; an $\alpha$ greater than 0.5 may favor learning skeleta \cite{cldice}. The clDice method relies on extracting accurate skeletons, and as such the performance of the method depends on hyperparameter tuning for an optimal \emph{k} value for a given dataset, which if done manually can be onerous and likely to find a suboptimal solution \cite{shamsi2023}. Additionally, the iterative process of min- and max- pooling used during soft skeletonization may result in loss of features that are critical to the topology of the structure.

\subsubsection{Homotopy Warping} The homotopic warping algorithm functions on the principle that given two binary masks with the same topology, one mask can be warped into the other mask by sequential flipping of simple points \cite{hu2022structureaware}. The connectivity of a pixel \emph{p} on a 2D binary image, or a voxel \emph{v} on a 3D binary volume, is defined by its neighboring pixels/voxels. The homotopy warping algorithm identifies topological critical points on a given binary mask (e.g., segmentation prediction) by warping it into another given binary mask (e.g., segmentation ground truth), and a resultant mask with the topological critical pixels, i.e., non-simple points, is used to compute a homotopy warping error-based loss where the critical pixels denote topological errors in warping one mask to the other. A distance transform is proposed by Hu \cite{hu2022structureaware} for optimal pixel flipping, and thus identification of critical pixels. Compared to clDice, the homotopic warping algorithm does not appear to increase Dice or adjusted Rand Index (ARI) scores as much as would be desired given the increase in training time \cite{hu2022structureaware}. However, the distance-ordered homotopy warping method can be used outside of the context of the homotopic warping loss function, and as such is useful as a generalizable method for identification of topologically critical points.


\section{Proposed Method}
\subsection{Morphological Smoothing using Mean Pixel Radius and Average Pooling}
As mentioned our research aim is to minimize a topology-preserving loss to precisely and accurately segment axons from 3D brain imagery, and find the centerlines of their curvilinear structures for automated annotation. As such we decided to formulate a loss function for this study. Our proposed method automatically determines the number of iterations required for a smoothing algorithm using geometric assessment of input data. The goal is to prevent over-thinning and loss of fine/small curvilinear structure features.

\subsubsection{Mean Pixel Radius for N Random Slices}
A tubular structure can be represented by its boundaries and the cross-sectional radius of each point in its skeleton \cite{DDT, GRELARD201789, kerautret:hal-01139374}, and so the maximum radius is the greatest distance from the medial axis of a skeleton point. For soft-skeletonization, using a \emph{k} parameter smaller than the largest pixel radius of a tubular structure in a dataset may result in incomplete thinning \cite{cldice}. We propose that for morphological approaches that rely on knowing the pixel radii of tubular structures in a given volumetric input, we may necessarily only need to know the maximum radius in a volumetric input with multiple tubular structures of varying cross-sectional radii. The maximum pixel radii of tubular structures from \emph{N} random slices of a given volumetric input can be averaged, and the result from that computation can be used as the number of iterations required to perform erosion and dilation operations for a thinning or smoothing algorithm. 

\begin{algorithm}
  \caption{Mean Pixel Radius}\label{tube_dist}
  \begin{algorithmic}[1]
    \Procedure{MPR}{$I$}\Comment{I is a 3D, 4D, or 5D input}
      \State $d \gets 0$
      \State $s \gets Slice(I)$
      \For{\texttt{slice in s}}
        \State $c \gets Canny(slice)$
        \State $Dist \gets MedialAxis(c)$
        \State $MaxDist \gets max(Dist)$
        \State $d \gets d + (2 \times MaxDist) $
      \EndFor
      \State \textbf{return} $int(2 \times (\frac{d}{len(s)}))$\Comment{of N random slices}
    \EndProcedure
  \end{algorithmic}
\end{algorithm}
\begin{algorithm}
  \caption{Topological Smoothing}\label{skel}
  \begin{algorithmic}[1]
    \Procedure{TS}{$I, k$}\Comment{k is predetermined using \textbf{MPR}}
      \State $I' \gets avepool(avepool(I))$
      \State $S \gets ReLU(I - I')$
      \For{\texttt{i to k}}
        \State $I \gets avepool(I)$
        \State $I' \gets avepool(avepool(I))$
        \State $Delta \gets ReLU(I-I')$
        \State $S \gets S + ReLU(Delta - S \circ Delta) $
      \EndFor
      \State \textbf{return} $S$
    \EndProcedure
  \end{algorithmic}
\end{algorithm}

Given a binary image \emph{I}, let \emph{d} be the mean pixel radius, initialized as zero. The \emph{Slice} function reshapes the 3D input and chooses without replacement, \emph{N} random slices, forming the resultant list of 2D inputs, \emph{s}. The boundary of tubular structures is determined using a Canny edge detector to enable computing the medial axis of the structure, and the distance to the edge boundary from the medial axis is returned by the \emph{MedialAxis} function from scikit-image. We experimentally found that returning twice the maximum distance and twice the mean pixel radius was more robust with our random sampling approach. Thus, the MPR method, Algorithm \ref{tube_dist}, returns a geometrically determined integer for estimating the iterations of thinning or smoothing for a given volumetric input with varying cross-sectional widths of tubular structures. 

\subsubsection{Topological Smoothing}
Some skeletonization algorithms produce constructs that are sensitive to rotations due to directional bias \cite{xie2003topology, DDT, steffensen1995method}, so we decided not to use a skeletonization approach for topology-preservation. Some of our initial experiments show that soft-skeletonization appears sensitive to data-dependant rotation transformations, likely due to increased morphological thinning at some rotations relative to others. We include these initial experiments in the Supplementary Material, but they form the basis for why we chose to use morphological smoothing. Following the logic of soft-skeletonization, we instead apply average pooling operations to the intermediate constructs of the algorithm, where we aim to induce topological smoothing by iteratively subtracting a more open construct from a less open construct; see line 7 in seen in Algorithm \ref{skel}. Using this method, we aim to produce an output that is less prone to over-thinning. For Algorithm \ref{skel}, \emph{k} is the number of iterations to perform the smoothing algorithm. We evaluate our approach on its ability to maintain geometrical connectivity as measured by metrics indicating topological preservation.

\subsection{Loss Function Formulations}
Similarly to clDice, we use the harmonic mean of the overlap of the masks and smoothed outputs,

\begin{equation} \label{eq5}
GATS(V_P, V_L) = 2 \times \frac{Tprec(T_P, V_L) \times Tsens(T_L, V_P)}{Tprec(T_P, V_L) + Tsens(T_L, V_P)} 
\end{equation}

where \emph{$T_P$} is the predicted smoothed output, and \emph{$V_L$} is the ground truth volume; similarly,  \emph{$T_L$} is the ground truth smoothed output, and  \emph{$V_P$} is the predicted segmentation mask. We believe that the overlap of smoothed output and ground truth still gives a relative measure of topological precision and topological sensitivity.  We used the  objective function in Eq. \ref{eq6} for training models with GATS, where $\alpha$ was fixed at 0.5. We formulated variants of the GATS loss function to systematically determine the effect of each component: 1) the MPR algorithm, 2) the smoothing algorithm, and 3) GATS.  We reasoned that if we use the MPR algorithm with min-, max- pooling we can still achieve soft skeletonization of the inputs, and so we compared  using the MPR method with both min-, max- pooling which enables soft skeletonization (GASK), and average pooling (GATS). We did an ablation study using our different loss functions.

\begin{equation}\label{eq6}
   L_{G} = (1 - \alpha)(1-\mathbf{Dice}) + \alpha(1-\mathbf{GATS})
\end{equation}

\setlength{\tabcolsep}{2pt}
\npdecimalsign{.}
\nprounddigits{3}

\begin{table*}[!htpb]
    \caption{\textbf{Mean and standard deviation evaluated on the test set of Dataset 1 (DS1) and Dataset 2 (DS2) across 10 trials. For clDice: $\alpha=0.65, k=3$, GASK and GATS both use MPR, for which N=10, GASK uses soft-skeletonization, GATS uses topological smoothing (TS). For TS: k=5 for DS1, k=3 for DS2.} } 
    \label{tab:pvgpe_table}
    \small
    \centering
    \begin{NiceTabular}{ccc  |c | c | c | c | c | c }
    \hline
        \textbf{Dataset} & \textbf{Attention} & \thead{\textbf{Model}} & \textbf{Dice $\uparrow$} & \textbf{clDice $\uparrow$} & \textbf{$\rho$-Dice $\uparrow$} & \textbf{Adj. Rand $\uparrow$} & \textbf{$\beta_{0}$ Error $\downarrow$} & \textbf{$\beta_{1}$ Error $\downarrow$} \\ \hline \hline
        ~ & ~ & \textit{clDice}  & 0.793   \text{\small$\pm$}  0.007 & 0.744   \text{\small$\pm$}  0.022 & 0.737   \text{\small$\pm$}  0.017 & 0.584   \text{\small$\pm$}  0.014 & 0.348   \text{\small$\pm$}  0.018 & 0.066   \text{\small$\pm$}  0.008 \\ \cline{3-9}
        
        DS1 & ~ None & \textit{ GASK} & 0.790   \text{\small$\pm$}  0.008 & 0.710   \text{\small$\pm$}  0.029 & 0.725   \text{\small$\pm$}  0.038 & 0.580   \text{\small$\pm$}  0.017 & 0.291   \text{\small$\pm$}  0.033 & 0.035   \text{\small$\pm$}  0.011 \\ \cline{3-9}
        
        ~ & ~ & \textit{ GATS} & 0.800   \text{\small$\pm$}  0.007 & 0.757   \text{\small$\pm$}  0.016 & 0.761   \text{\small$\pm$}  0.013 & 0.598   \text{\small$\pm$}  0.014 & 0.316   \text{\small$\pm$}  0.019 & 0.061   \text{\small$\pm$}  0.008 \\ \cline{2-9}
                

        ~ & ~ & \textit{clDice}  & 0.801   \text{\small$\pm$}  0.011 & 0.761   \text{\small$\pm$}  0.026 & 0.759   \text{\small$\pm$}  0.025 & 0.600   \text{\small$\pm$}  0.022 & 0.334   \text{\small$\pm$}  0.016 & 0.074   \text{\small$\pm$}  0.014 \\ \cline{3-9}
        
        ~ & ~ SCA & \textit{ GASK } & 0.790   \text{\small$\pm$}  0.012 & 0.701   \text{\small$\pm$}  0.027 & 0.718   \text{\small$\pm$}  0.031 & 0.579   \text{\small$\pm$}  0.024 & \bftab{0.270   \text{\small$\pm$}  0.039} & 0.074   \text{\small$\pm$}  0.013 \\ \cline{3-9}
        
        ~ & ~ & \textit{ GATS } & \bftab{0.819   \text{\small$\pm$}  0.006} & \bftab{0.799   \text{\small$\pm$}  0.012} & \bftab{0.798   \text{\small$\pm$}  0.009} & \bftab{0.636   \text{\small$\pm$}  0.011} & 0.314   \text{\small$\pm$}  0.022 & \bftab{0.060   \text{\small$\pm$}  0.007} \\ \hline \hline

         ~ & ~ & \textit{clDice}  & 0.788   \text{\small$\pm$}  0.014 & 0.689   \text{\small$\pm$}  0.027 & 0.775   \text{\small$\pm$}  0.029 & 0.576   \text{\small$\pm$}  0.027 & 0.127   \text{\small$\pm$}  0.031 & 0.151   \text{\small$\pm$}  0.017 \\ \cline{3-9}
        
        DS2 & ~ None & \textit{ GASK } & 0.698   \text{\small$\pm$}  0.018 & 0.741   \text{\small$\pm$}  0.013 & 0.804   \text{\small$\pm$}  0.011 & 0.395   \text{\small$\pm$}  0.036 & 0.024   \text{\small$\pm$}  0.009 & 0.370   \text{\small$\pm$}  0.017 \\ \cline{3-9}
        
        ~ & ~ & \textit{ GATS } & \bftab{0.803   \text{\small$\pm$}  0.007} & 0.715   \text{\small$\pm$}  0.016 & \bftab{0.806   \text{\small$\pm$}  0.016} & \bftab{0.605   \text{\small$\pm$}  0.014} & 0.047   \text{\small$\pm$}  0.010 & \bftab{0.096   \text{\small$\pm$}  0.014} \\ \cline{2-9}
        
        
                

        ~ & ~ & \textit{clDice}  & 0.725   \text{\small$\pm$}  0.008 & \textbf{0.746   \text{\small$\pm$}  0.008} & 0.799   \text{\small$\pm$}  0.007 & 0.450   \text{\small$\pm$}  0.016 & 0.091   \text{\small$\pm$}  0.023 & 0.367   \text{\small$\pm$}  0.016 \\ \cline{3-9}
        
        ~ & ~ SCA & \textit{ GASK } & 0.691   \text{\small$\pm$}  0.019 & 0.738   \text{\small$\pm$}  0.025 & 0.796   \text{\small$\pm$}  0.023 & 0.381   \text{\small$\pm$}  0.038 & \bftab{0.021   \text{\small$\pm$}  0.004} & 0.374   \text{\small$\pm$}  0.008 \\ \cline{3-9}
        
        ~ & ~ & \textit{ GATS } & 0.787   \text{\small$\pm$}  0.011 & 0.682  \text{\small$\pm$}  0.021 & 0.772   \text{\small$\pm$}  0.024 & 0.573   \text{\small$\pm$}  0.021 & 0.050   \text{\small$\pm$}  0.012 & 0.104   \text{\small$\pm$}  0.021 \\
        \hline
    \end{NiceTabular}
\end{table*}

\setlength{\tabcolsep}{2pt}
\npdecimalsign{.}
\nprounddigits{2}

\begin{table*}[!htpb]
    \caption{\textbf{Slicing experiments for MPR (Algorithm \ref{tube_dist}). Mean and Standard Deviation of Evaluation Metrics for Differently Sliced Variants of GATS on the Test Set of Dataset 1 and Dataset 2 across 10 trials, For clDice: $\alpha=0.65, k=3$. For TS: k=5 for DS1, k=3 for DS2. Time indicates average minutes for a single trial (n=10).}} 
    \label{tab:time_table}
    \centering
    \small
     \begin{NiceTabular}{cc  |c | c | c | c | c | c | c }
    \hline
        \textbf{Dataset} & \thead{\textbf{Model}} & \textbf{Dice $\uparrow$} & \textbf{clDice $\uparrow$} & \textbf{$\rho$-Dice $\uparrow$} & \textbf{Adj. Rand $\uparrow$} & \textbf{$\beta_{0}$ Error $\downarrow$} & \textbf{$\beta_{1}$ Error $\downarrow$} & \textbf{Time $\downarrow$} \\ \hline \hline
        
        

        DS1 & clDice & 0.793   \text{\small\text{\small$\pm$}}  0.007 & 0.744   \text{\small\text{\small$\pm$}}  0.022 & 0.737   \text{\small\text{\small$\pm$}}  0.017 & 0.584   \text{\small\text{\small$\pm$}}  0.014 & 0.348   \text{\small\text{\small$\pm$}}  0.018 & 0.066   \text{\small\text{\small$\pm$}}  0.008  & \bftab{26.627 \text{\small$\pm$} 7.676} \\ \cline{2-9}
        
        ~ & \textit{GATS N=2} & 0.803  \text{\small$\pm$}  0.006 & 0.756  \text{\small$\pm$}  0.014 & 0.762  \text{\small$\pm$}  0.015 & 0.605  \text{\small$\pm$}  0.012 & 0.339  \text{\small$\pm$}  0.010 & 0.066  \text{\small$\pm$}  0.003  & {38.507 \text{\small$\pm$} 11.088} \\ \cline{3-9}
        
        ~ & \textit{GATS N=3} & 0.803  \text{\small$\pm$}  0.007 & 0.761  \text{\small$\pm$}  0.014 & 0.767  \text{\small$\pm$}  0.015 & 0.605  \text{\small$\pm$}  0.013 & \bftab{0.321  \text{\small$\pm$}  0.021} & \bftab{0.058  \text{\small$\pm$}  0.007}  & {46.307 \text{\small$\pm$} 13.314} \\ \cline{3-9}

        ~ & \textit{GATS N=4} & \bftab{0.808  \text{\small$\pm$}  0.004} & \bftab{0.773  \text{\small$\pm$}  0.009} & \bftab{0.778  \text{\small$\pm$}  0.010} & \bftab{0.615  \text{\small$\pm$}  0.008} & 0.334  \text{\small$\pm$}  0.024 & 0.061  \text{\small$\pm$}  0.006  & {38.398 \text{\small$\pm$} 11.050} \\ \cline{3-9} 
        
        ~ & \textit{GATS N=5} & 0.803  \text{\small$\pm$}  0.006 & 0.765  \text{\small$\pm$}  0.014 & 0.768  \text{\small$\pm$}  0.017 & 0.604  \text{\small$\pm$}  0.012 & 0.331  \text{\small$\pm$}  0.017 & 0.066  \text{\small$\pm$}  0.009  & {42.945 \text{\small$\pm$} 12.348} \\ \hline \hline
        
        DS2 & clDice & 0.788   \text{\small\text{\small$\pm$}}  0.014 & 0.689   \text{\small\text{\small$\pm$}}  0.027 & 0.775   \text{\small\text{\small$\pm$}}  0.030 & 0.576   \text{\small\text{\small$\pm$}}  0.027 & 0.127   \text{\small\text{\small$\pm$}}  0.031 & 0.151   \text{\small\text{\small$\pm$}}  0.017  & \bftab{30.723 \text{\small$\pm$} 8.845} \\ \cline{2-9}

        ~ & \textit{GATS N=2} & 0.795  \text{\small$\pm$}  0.007 & 0.700  \text{\small$\pm$}  0.016 & 0.789  \text{\small$\pm$}  0.015 & 0.589  \text{\small$\pm$}  0.016 & \bftab{0.045  \text{\small$\pm$}  0.010} & \bftab{0.094  \text{\small$\pm$}  0.012}  & {34.483 \text{\small$\pm$} 9.939} \\ \cline{3-9}

        ~ & \textit{GATS N=3} & 0.796  \text{\small$\pm$}  0.006 & 0.701  \text{\small$\pm$}  0.015 & 0.793  \text{\small$\pm$}  0.013 & 0.591  \text{\small$\pm$}  0.011 & 0.052  \text{\small$\pm$}  0.009 & 0.102  \text{\small$\pm$}  0.012  & {35.299 \text{\small$\pm$} 10.168} \\ \cline{3-9}
        
        ~ & \textit{GATS N=4} & 0.800  \text{\small$\pm$}  0.008 & \bftab{0.710  \text{\small$\pm$}  0.021} & \bftab{0.801  \text{\small$\pm$}  0.020} & 0.599  \text{\small$\pm$}  0.016 & 0.049  \text{\small$\pm$}  0.009 & 0.101  \text{\small$\pm$}  0.015  & {36.480 \text{\small$\pm$} 10.512} \\ \cline{3-9}

        ~ & \textit{GATS N=5} & \bftab{0.800  \text{\small$\pm$}  0.006} & 0.709  \text{\small$\pm$}  0.014 & 0.800  \text{\small$\pm$}  0.013 & \bftab{0.599  \text{\small$\pm$}  0.013} & 0.051  \text{\small$\pm$}  0.005 & 0.106  \text{\small$\pm$}  0.009  & {34.787 \text{\small$\pm$} 10.018} \\  \hline

    \end{NiceTabular}
\end{table*}

\begin{table*}[!htpb]
\caption{\textbf{Ablation study mean and standard deviation evaluated on the test set of Dataset 1 (DS1) and Dataset 2 (DS2) across 10 trials. MPR is Mean Pixel Radius,  Smooth is Topological Smoothing, Soft Skel is soft-skeletonization}} 
\label{tab:ab_table}
\centering
\small
\begin{NiceTabular}{cccc | c | c | c | c | c | c }
\hline
\multicolumn{2}{c}{\textbf{Ablated}} & \textbf{Dataset} & \textbf{Model} & \textbf{Dice $\uparrow$} & \textbf{clDice $\uparrow$} & \textbf{$\rho$-Dice $\uparrow$} & \textbf{Adj. Rand $\uparrow$} & \textbf{$\beta_{0}$ Error $\downarrow$} & \textbf{$\beta_{1}$ Error $\downarrow$} \\ \hline \hline
\multicolumn{2}{l}{\multirow{4}{*}{MPR}} &{\multirow{2}{*}{DS1}}                ~ &  \textit{Soft Skel (k=5)} &  0.794 \text{\small$\pm$} 0.010 & 0.707 \text{\small$\pm$} 0.023 & 0.718 \text{\small$\pm$} 0.031 & 0.586 \text{\small$\pm$} 0.020 & 0.253 \text{\small$\pm$} 0.058 & 0.070 \text{\small$\pm$} 0.015 \\ 
\multicolumn{2}{c}{}                   &                                        ~ &   \textit{Smooth (k=5)} &  0.810 \text{\small$\pm$} 0.009 & 0.782 \text{\small$\pm$} 0.017 & 0.781 \text{\small$\pm$} 0.015 & 0.618 \text{\small$\pm$} 0.018 & 0.326 \text{\small$\pm$} 0.036 & 0.065 \text{\small$\pm$} 0.010    \\ \cline{4-10}
\multicolumn{2}{c}{}                   & \multirow{2}{*}{DS2}                   ~ &  \textit{Soft Skel (k=5)} & 0.749 \text{\small$\pm$} 0.008 & 0.715 \text{\small$\pm$} 0.019 & 0.801 \text{\small$\pm$} 0.006 & 0.497 \text{\small$\pm$} 0.017 & 0.095 \text{\small$\pm$} 0.038 & 0.303 \text{\small$\pm$} 0.020    \\ 
\multicolumn{2}{c}{}                   &                                         ~ &   \textit{Smooth (k=5)} &  0.801 \text{\small$\pm$} 0.010 & 0.710 \text{\small$\pm$} 0.019 & 0.794 \text{\small$\pm$} 0.018 & 0.601 \text{\small$\pm$} 0.019 & 0.053 \text{\small$\pm$} 0.010 & 0.102 \text{\small$\pm$} 0.019    \\ \hline \hline
\multicolumn{2}{l}{\multirow{4}{*}{Smooth}} & \multirow{2}{*}{DS1}                   ~ & \textit{No Attn} &  0.804 \text{\small$\pm$} 0.007 & 0.756 \text{\small$\pm$} 0.012 & 0.768 \text{\small$\pm$} 0.015 & 0.607 \text{\small$\pm$} 0.014 & 0.262 \text{\small$\pm$} 0.066 & 0.057 \text{\small$\pm$} 0.016    \\
\multicolumn{2}{c}{}                   &                                          ~ & \textit{SCA} &  0.798 \text{\small$\pm$} 0.032 & 0.734 \text{\small$\pm$} 0.089 & 0.740 \text{\small$\pm$} 0.090 & 0.594 \text{\small$\pm$} 0.064 & 0.187 \text{\small$\pm$} 0.074 & 0.046 \text{\small$\pm$} 0.016 \\ \cline{4-10}
\multicolumn{2}{c}{}                   & \multirow{2}{*}{DS2}                     ~ & \textit{No Attn} &  0.812 \text{\small$\pm$} 0.001 & 0.735 \text{\small$\pm$} 0.001 & 0.826 \text{\small$\pm$} 0.001 & 0.623 \text{\small$\pm$} 0.001 & 0.083 \text{\small$\pm$} 0.005 & 0.151 \text{\small$\pm$} 0.004    \\ 
\multicolumn{2}{c}{}                   &                                          ~ & \textit{SCA} & 0.792 \text{\small$\pm$} 0.060 & 0.695 \text{\small$\pm$} 0.012 & 0.785 \text{\small$\pm$} 0.014 & 0.584 \text{\small$\pm$} 0.012 & 0.095 \text{\small$\pm$} 0.009 & 0.147 \text{\small$\pm$} 0.008    \\ \hline \hline
GATS                                                                            ~ &  ~ & DS1 &   \textit{Dice} &    0.805 \text{\small$\pm$} 0.011 & 0.763 \text{\small$\pm$} 0.038 & 0.769 \text{\small$\pm$} 0.038 & 0.609 \text{\small$\pm$} 0.023 & 0.196 \text{\small$\pm$} 0.066 & 0.048 \text{\small$\pm$} 0.020    \\ 
                                                                                ~ & ~ & DS2 &   \textit{Dice}   &   0.776 \text{\small$\pm$} 0.002 & 0.663 \text{\small$\pm$} 0.003 & 0.751 \text{\small$\pm$} 0.004 & 0.551 \text{\small$\pm$} 0.003 & 0.076 \text{\small$\pm$} 0.011 & 0.119 \text{\small$\pm$} 0.008    \\   \hline 
\end{NiceTabular}
\end{table*}

\begin{table}[!htpb]
\caption{\textbf{Betti error mean and standard deviation evaluated on the test set of Janelia across 10 trials. All models use SCA attention. For clDice: $\alpha=0.65, k=3$. For Smooth Dice and GATS: N=4.}} 
\label{tab:jan_table}
\centering
\small
\begin{NiceTabular}{cc|c|c}
\textbf{Architecture} &\textbf{Model} & \textbf{$\beta_{0}$ Error $\downarrow$} & \textbf{$\beta_{1}$ Error $\downarrow$} \\ \hline \hline
\multirow{5}{*}{3DResSE UNet} & \textit{clDice}   & 0.614 ± 0.159 & 0.047 ± 0.005   \\ \cline{3-4}
                              & \textit{Warping}   & 0.716 ± 0.003 & 0.049 ± 0.001   \\ \cline{3-4}
                              & \textit{Dice}   & 0.685 ± 0.005 & 0.052 ± 0.002    \\ \cline{3-4}
                              & \textit{Smooth Dice} & 0.709 ± 0.178 & 0.048 ± 0.013    \\ \cline{3-4}
                              & \textit{GATS}   & \textbf{0.200 ± 0.002} & \textbf{0.036 ± 0.003}    \\ \hline \hline
\multirow{5}{*}{Cascading 3D UNet} & \textit{clDice} & 0.641 ± 0.004 & 0.050 ± 0.001    \\ \cline{3-4}
                              & \textit{Warping}   & 0.600 ± 0.001 & \textbf{0.049 ± 0.001}    \\ \cline{3-4}
                              & \textit{Dice}   & 0.700 ± 0.012 & 0.050 ± 0.006    \\ \cline{3-4}
                              & \textit{Smooth Dice} & \textbf{0.394 ± 0.177} & 0.052 ± 0.001    \\ \cline{3-4}
                              & \textit{GATS}   & 0.631 ± 0.001 & 0.051 ± 0.001    \\ \hline 
\end{NiceTabular}
\end{table}

\begin{figure}
    \centering
    \includegraphics[width=0.48\textwidth]{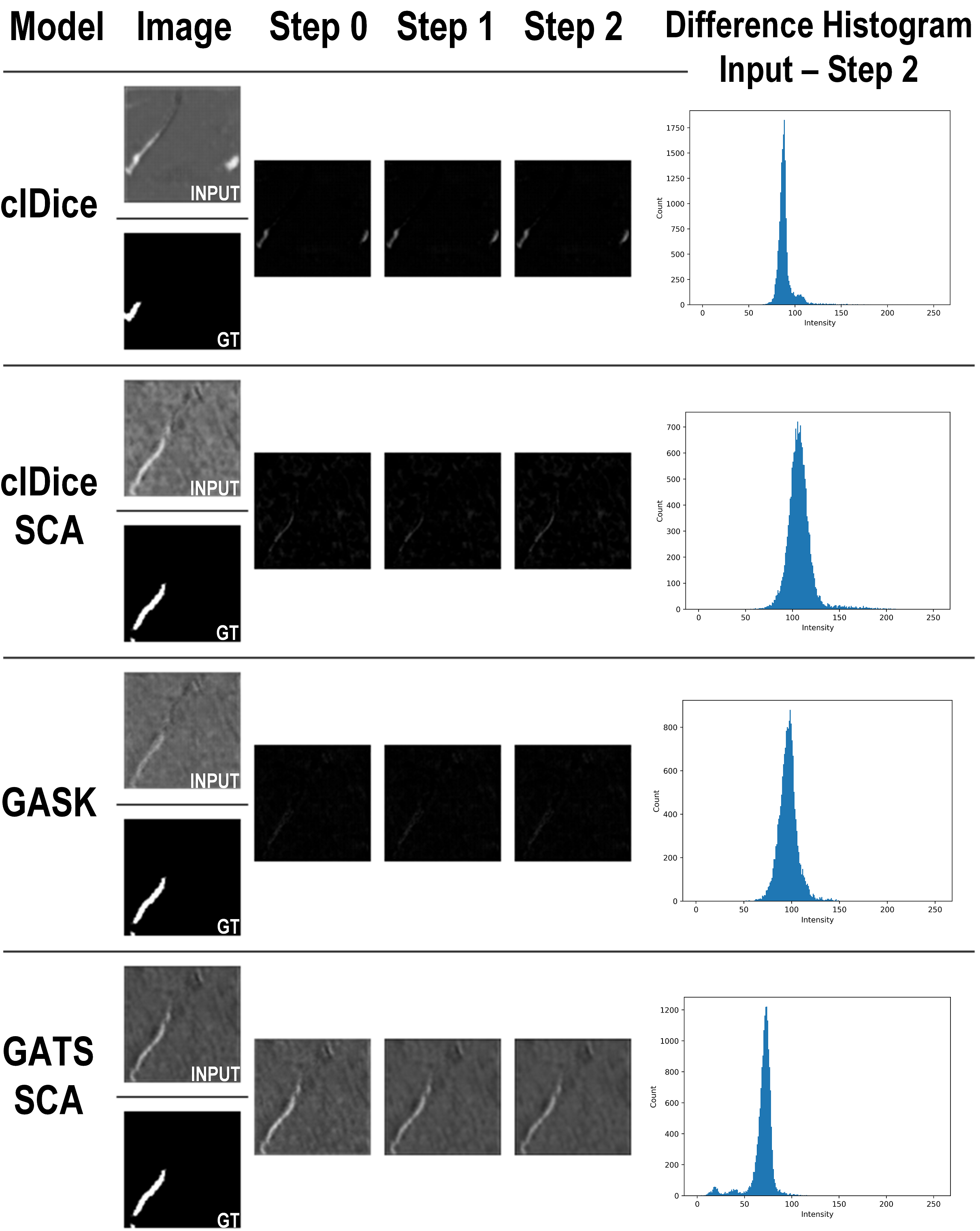}
    \caption{\textbf{Morphological thinning progress on DS2.} The best performing models on DS2 are compared. Iterations of soft-skeletonization (GASK) or topological smoothing (GATS), which are automatically determined by MPR while manually set for clDice, are both three. Therefore, three steps of skeletonization/smoothing are shown. The difference histogram between the input and Step 2 shows  difference in pixel intensities, where the smoothed output difference results in intensities centered around lower ranges, indicating that the input and Step 2 output are closer in pixel intensities. This is expected for a smoothing operation. SCA: spatial and channel attention. GT: ground truth.}
    \label{fig:fig1}
\end{figure}

\section{Experimental Methodology}
\subsection{Datasets}
As we are interested in automated annotation of unannotated volumes of axons, we conducted our experiments using 3D brain imagery data.

\subsubsection{Dataset 1} Dataset 1 (DS1) is a light sheet microscopy dataset of 3X expanded mouse brain tissue, where a stain was used to target Parvalbumin positive neurons from the globus pallidus externus (PVGPe). The CLARITY method was used to stabilize the tissue with clear hydrogels, preserving biomolecules and enabling removal of lipids, which makes the unstained portions of sample optically transparent \cite{clarity}. The PVGPe volume in its entirety is 2048×2048×1271 voxels, and has a voxel resolution of 0.6×0.6×2 µm, where only a 256×256×206 voxel (148×148×412 µm) subvolume was manually annotated \cite{pollack2022axon}. Model training was done on 128×128×64 sized voxel samples using a contiguous subdivided volumes of DS1 in a 50:25:25 training, test and validation split.

\subsubsection{Dataset 2} Dataset 2 (DS2) consists of 20X magnified samples from the mouse thalamus which were labeled via cortical injection with recombinant adeno-associated virus expressing tdTomato (red) and synaptophysin (green). Imagery was then acquired using a Leica confocal microscope. The tdTomato channel was converted to grayscale for use in this paper. The cross-section of this data volume is 581.250 $\mu m^2$, thickness is 35 µm, lateral pixel resolution is 0.142 $\mu m^2$, and axial resolution is 0.69 µm. The full DS2 volume is 4096×4096×52 voxels. The training, testing, and validation volumes were split in the same ratios as those for DS1, except the voxel size samples were 128×128×32 for model training.

\subsubsection{Janelia} 
We also conducted experiments using the Janelia dataset from the BigNeuron Project \cite{peng2015bigneuron}, which consists of optical microscopy data of single neurons from the adult Drosophila nervous system. Janelia has 42 volumes of data, and so we allocated 30 volumes for training, six for validation and six for testing. We ended up using only one volume for testing in case more training data was needed. All volumes were scaled between zero and one using min-max normalization and we used crop size of 128×128×32 for model training.

\subsection{Model Implementation and Training}
Our experiments on DS1 and DS2 consisted of training a Residual 3D UNet with four resolution blocks using one of our formulated loss functions and comparing performance against clDice \cite{cldice}. Our experiments on Janelia consisted of using a Residual 3D UNet with squeeze and excitation blocks (3DResSE UNet), and a Cascading 3D UNet as previously described \cite{pollack2022axon} with four and three resolution blocks for the voxel-wise segmentation head and centerline detection head, respectively. The Cascading 3D UNet used a multi-input loss formulated as:

\begin{equation}\label{eq7}
   L_{M} = (1 - \alpha)(1-\mathbf{seg loss}) + \alpha(1-\mathbf{cl loss})
\end{equation}

where $\alpha$=0.8, and \emph{seg loss} (segmentation loss) was the chosen loss function for the voxel-wise segmentation task, and \emph{cl loss} (centerline loss) was the chosen loss function for the centerline prediction task. We also compared performance on the Janelia dataset against Warping loss \cite{hu2022structureaware}. Warping loss supersedes both TopoNet loss \cite{toponet} and DMT loss \cite{dmt}, and so we only tested against Warping loss. We tried GATS with N=2, 3, 4, 5, or 10 while profiling the average training time required for the 10 trials. Optimal hyperparameters for clDice were determined experimentally by testing $\alpha$=0.5 or 0.65, and \emph{k}=3 or 8. All our models were trained using an Intel Xeon G6 node (40 cores) with 2 NVIDIA Volta V100 GPUs of 337 GiB memory. 

We used a PyTorch framework for data processing, algorithm development and model training. All data used were pre-processed by clipping the highest and lowest 0.01\% of values, applying a median filter, and scaling between 0 and 1. Besides using data augmentation during training and inference as previously described in \cite{pollack2022axon}, we used a series of affine rotation transformations that were particular for each dataset. Each experiment was repeated 10 times using a model with or without 3D spatial and channel attention (SCA) \cite{SCA}. We tried efficient channel attention \cite{ECA} and triplet attention \cite{triplet} as well (Supplementary Material), but found that evaluation metrics were lower than those for models trained on SCA, so only SCA results are reported. 

The 3D U-Net implementation consisted of a 3×3×3 convolution layer followed by group normalization and activation using exponential linear units. Strided 2×2×2 max-pooling and strided transpose convolutions with max pooling were used for downsampling and upsampling, respectively. An ADAM optimizer coupled with cosine annealing was used, with an initial learning rate of 1×$10^{-4}$ and weight decay of 1×$10^{-3}$. Each trial consisted of randomly cropped and augmented samples from the input data in mini batches of 16 samples. The segmentation output from each model was skeletonized using a 3D-skeletonization algorithm to acquire single voxel wide centerlines. Models were evaluated using the Dice coefficient for voxel-wise similarity \cite{zou2004statistical}, the clDice metric \cite{cldice} and Betti number errors for topology preservation \cite{Robins2002}, $\rho$-Dice coefficient for centerline detection accuracy \cite{pollack2022axon}, and the adjusted Rand Index (ARI) for ground truth and predicted clusterings equivalence \cite{robert2021comparing}. The Betti numbers $\beta_0$ and $\beta_1$ measure the number of distinct connected components and circular holes, respectively \cite{cldice}. All tables bold the best metric in each category.

\begin{figure}
    \centering
    \includegraphics[width=0.48\textwidth]{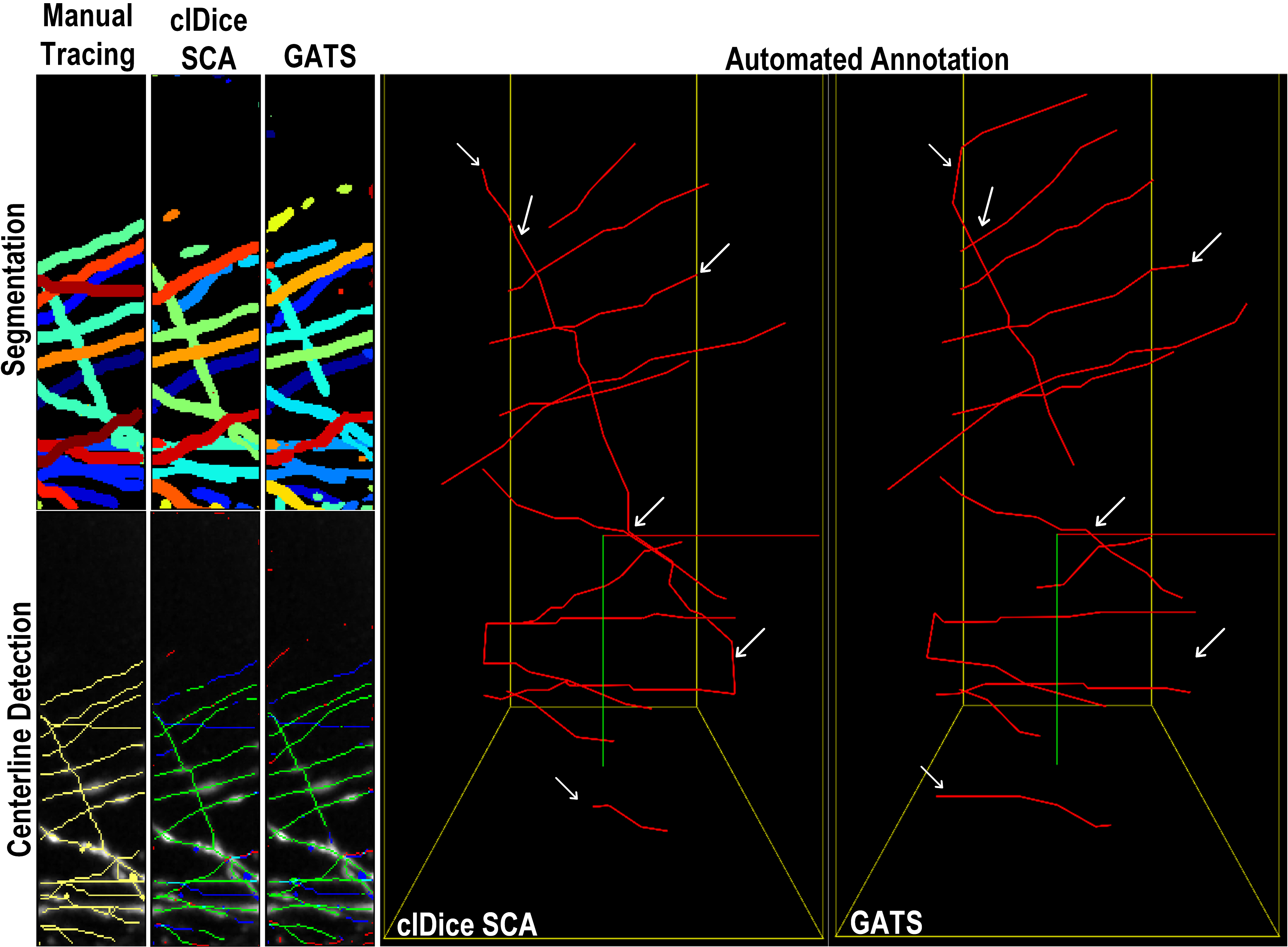}
    \caption{\textbf{Comparison of 3D Axon Projections for DS1.} The segmentation results (top) are contrasted with the centerline detection results (bottom), with automated annotation of a 3D volume (right). For centerline detection results, true positives (\textcolor{green}{green}), false negatives (\textcolor{blue}{blue}), and false positives (\textcolor{red}{red}) are shown. The red label for false positive does not apply to the automated annotation results. The white arrows show differences in axonal connectivity across models.}
    \label{fig:fig2}
\end{figure}

\section{Results}
\subsection{Axon Segmentation and Centerline Detection}
We compared GATS against two topology-preserving losses, one which uses skeletonization (clDice) and one which does not (Warping). The performance of Warping loss was only compared for the Janelia dataset as even after 24 hours Warping loss does not finish training three of 10 trials on DS1 or DS2, which does not make a reasonable comparison for either clDice or our method.

For results in Table \ref{tab:pvgpe_table}, MPR algorithmically determined its own number of iterations to perform either soft skeletonization (GASK) or topological smoothing (GATS). A comparison of thinning or smoothing outputs is shown (Figure \ref{fig:fig1}). MPR (Algorithm \ref{tube_dist}) determined \emph{k}=5 for DS1 and \emph{k}=3 for DS2. For DS1 we found that when the number of iterations used to perform skeletonization are acquired via the MPR method, the model performance drops unless topological smoothing is performed (GATS), as seen in Table \ref{tab:pvgpe_table} and Figure \ref{fig:fig2}, indicating the need for morphological smoothing for topological preservation. The GASK models perform poorer on DS1 because MPR determined \emph{k}=5 leads to over-thinning with a method like soft-skeletonization, but not with the topological smoothing method. For DS2 GASK without attention is comparable in performance to clDice with manually selected hyperparameters, indicating the efficacy of using geometric assessment for determining the optimal number of iterations for morphological thinning (see Figure \ref{fig:fig3} for a visual comparison). On DS2 the GATS model performs better for all metrics except the centerline detection metric (clDice) than the model trained on the clDice loss as alpha is affixed to 0.5 for GASK/GATS models. This shows the sensitivity of alpha and k parameter choices for segmentation and centerline detection. Ideally, hyperparameter tuning can be data driven and not manually optimized, as GASK/GATS aims to do. The MPR algorithm can select any \emph{N} random slices (without replacement) from a given input volume, and the training time of N = 2, 3, 4, 5 and 10 was compared. We found that when N=10, the mean training time for 10 trials with GATS was double that of the mean training time of trials with clDice (5.2h vs. 10h). However, choosing N=4 seemed sufficient for each dataset, though slower by 1h for DS2 and 2h for DS1 (Table \ref{tab:time_table}); this is most likely due to the use of the Canny detector for boundary determination. Based on our evaluation, it would appear that adding an attention mechanism is useful for datasets with lower voxel resolution like DS1. 

\subsection{Ablation Study}
We carried out an ablation study by systematically removing one of three components from GATS in three separate trials using models without attention unless indicated otherwise. The first trial consisted of removing \emph{MPR} and manually setting \emph{k}=5 as it was previously determined by MPR as the optimal number of iterations for topological smoothing for DS1 but not DS2. The second trial consists of removing the average pooling-based morphological thinning and using a general thinning algorithm from Scikit Image to perform morphological thinning while using MPR to determine the number of iterations to perform. The third trial consisted of removing GATS and using a Dice loss. For DS1, GASK without attention is comparable to MPR-ablated Soft Skel (\emph{k}=5), where k=5 was the number of iterations previously determined by MPR as optimal for DS1. GATS without attention has a worse evaluation metrics than MPR-ablated Smooth(\emph{k}=5), though the latter has worse Betti error rates, which is unexpected but maybe likely due to the random sampling approach for MPR which gave an incorrect assessment for at least one of the ten trials used to compute the average for GATS without attention. However, given that MPR-ablated \emph{k}=5 model with topological smoothing (Smooth, \emph{k}=5) on DS1 performs better than clDice without attention (0.782 vs. 0.744), there is merit to finding the number of iterations required for a thinning or smoothing algorithm as trials in Table \ref{tab:pvgpe_table} use \emph{k}=3 for DS1. On DS2 the MPR-ablated model (\emph{k}=5) with topological smoothing performs marginally worse than its GATS counterpart without attention (Table \ref{tab:pvgpe_table}, row 9 vs. Table \ref{tab:ab_table}, row 4). The same holds for the MPR-ablated soft-skeletonization model without attention, though its Dice, ARI and Betti 1 scores are marginally better for the MPR ablated version. The attention model for DS1 without topological smoothing but with a general thinning algorithm for skeletonization has a worse centerline line detection metric (clDice) versus GATS with attention (0.735 vs. 0.799), indicating that for DS1 the average-pooling based topological smoothing was effective for preserving topological connectivity. However the smoothing-ablated model has marginally less Betti errors relative to both clDice and GATS, indicating the need to explore different approaches for morphological thinning or smoothing. For DS2, removing attention, topological smoothing and using a general thinning algorithm results in a more performant model  (Table \ref{tab:pvgpe_table}, row 9 vs. Table \ref{tab:ab_table}, row 7) relative to the dice metrics used for comparison. This indicates that for inputs with higher voxel resolution topological smoothing might be deleterious, while these inputs may still benefit from topological skeletonization (Table \ref{tab:pvgpe_table}, row 8 vs. Table \ref{tab:ab_table}, row 7).

\subsection{Comparing Topology-Preservation on Janelia}
For automated annotation of unannotated brain imagery, a simpler solution may be to improve curvilinear structure segmentation using a topology preserving loss without any multitasking for centerline detection. We compared the performance of topology-preserving losses via their Betti errors on two architectures: 1) 3DResSE UNet, and 2) a multi-headed Cascading 3D UNet as described earlier \cite{pollack2022axon}. MPR determined \emph{k}=3 for Janelia. In Table \ref{tab:jan_table}, Smooth Dice was given inputs that were passed through topological smoothing. Based on the best scores for the models trained on the loss functions we compared, we find that topological smoothing reduces Betti error rates for the Janelia dataset.

\subsection{Automated Annotation of Unannotated Brain Imagery}

\begin{figure}
    \centering
    \includegraphics[width=0.48\textwidth]{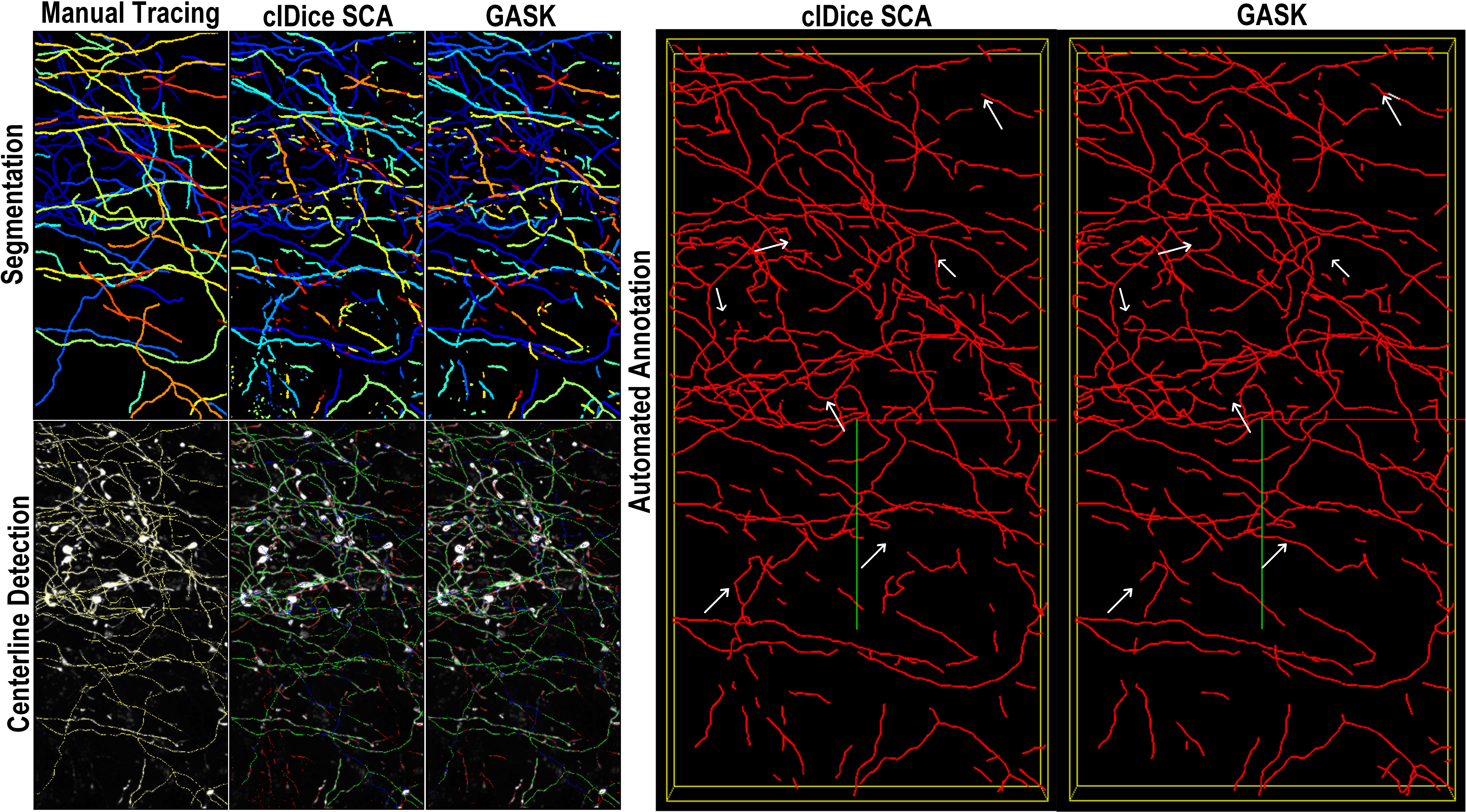}
    \caption{\textbf{Comparison of 3D Axon Projections for DS2.} 3D axon predictions of DS2 show that the GASK version produces longer and more connected axon segments (white arrows), which shows that soft-skeletonization can overthin if an incorrect \emph{k} hyperparameter is chosen. For centerline detection results, true positives (\textcolor{green}{green}), false negatives (\textcolor{blue}{blue}), and false positives (\textcolor{red}{red}) are shown. The red label for false positive does not apply to the automated annotation results.}
    \label{fig:fig3}
\end{figure}

To visualize automatic annotation of unannotated 3D brain imagery, we used NeuroTrALE \cite{neurotrale}, a variant of Neuroglancer, which is a WebGL-based viewer for volumetric data. We used the best performing clDice or GASK/GATS trained model according to the clDice metric on DS1 and DS2. As mentioned, Warping loss does not finish training even three of 10 trials on DS1 or DS2 in 24 hours, and so was not used as a comparison on these datasets. Models trained on our method produce longer and less disconnected axons, as seen in Figure \ref{fig:fig2} for DS1 and Figure \ref{fig:fig3} for DS2. For the centerline detection results in Figures \ref{fig:fig2} and \ref{fig:fig3}, green indicates true positive, blue indicates false negative and red indicates false positive for model predictions relative to the ground truth. Manual tracings are the ground truth. 

\section{Discussion}
 Topology-preserving methods which use skeletonization may over-thin 3D brain imagery, and so techniques such as average pooling-based smoothing, as seen in our topological smoothing method, may be needed. However, our ablation study suggests that the approach used by both soft-skeletonization and topological smoothing is possibly too aggressive for the type of 3D brain imagery used here, even though a loss function formulated using topological smoothing performs relatively better for segmentation and centerline detection. We find it limiting that our best models are all some flavor of a UNet architecture, and aim to incorporate our findings into other deep learning architectures, e.g., like an adapter module for Segment Anything \cite{MedSAM}. 
\section{Conclusive Remarks}
We show that the optimal number of thinning or smoothing iterations for a morphological operation can be determined using N random slices for a volumetric input of brain imagery, and that average pooling-based morphological smoothing can improve both segmentation and centerline detection metrics, indicating increased topological preservation. We also show that our loss function, GATS, can be used to train models for automatic annotation of volumes of brain imagery.

\section{Acknowledgement}
The authors acknowledge Kwanghun Chung at MIT for providing data, and MIT Lincoln Laboratory Supercomputing Center for their support
of high performance computing tasks. This work was supported by the National Institutes of Health (NIH) U01MH117072, NIMH 1R43MH128076-01,
and MBF Bioscience. Any opinions, conclusions, or recommendations
in this material are those of the author(s) and do
not necessarily reflect the views of NIH/NIMH.

{\small
\bibliographystyle{ieee_fullname}
\bibliography{egbib}
}

\end{document}